\documentclass[twocolumn,aps,prl,groupedaddress,nofootinbib,superscriptaddress,floatfix,amsmath,amssymb,preprintnumbers]{revtex4-1}
\usepackage{graphicx}
\usepackage{dcolumn}
\usepackage{bm}
\usepackage{slashed}
\usepackage{siunitx}
\usepackage{ulem,xpatch}
\usepackage{hyperref}
\usepackage[mathlines]{lineno}

\usepackage{comment}
\usepackage{soul}
\usepackage{xcolor}
\usepackage{xfrac}
\hypersetup{colorlinks, linkcolor = [rgb]{0,0.0,0.5}, citecolor = [rgb]{0,0.0,0.5}, urlcolor = [rgb]{0,0.0,0.5}}

\allowdisplaybreaks[1]

\begin{document}


\preprint{JLAB-THY-19-3057}
\preprint{SLAC-PUB-17477}

\title{Unified Description of Polarized and Unpolarized Quark Distributions in the Proton}

\newcommand*{\JLAB}{Thomas Jefferson National Accelerator Facility, Newport News, Virginia 23606, USA}\affiliation{\JLAB}
\newcommand*{\DUKE}{Department of Physics, Duke University, Durham, North Carolina 27708, USA}\affiliation{\DUKE}
\newcommand*{\COSTARICA}{Laboratorio de F\'isica Te\'orica y Computacional, Universidad de Costa Rica, 11501 San Jos\'e, Costa Rica}\affiliation{\COSTARICA}
\newcommand*{\HEIDELBERG}{Institut f\"ur Theoretische Physik der Universit\"at, D-69120 Heidelberg, Germany}\affiliation{\HEIDELBERG}
\newcommand*{\SLAC}{SLAC National Accelerator Laboratory, Stanford University, Stanford, California 94309, USA}\affiliation{\SLAC}

\author{Tianbo~Liu}\email{liutb@jlab.org}\affiliation{\JLAB}\affiliation{\DUKE}
\author{Raza~Sabbir~Sufian}\affiliation{\JLAB}
\author{Guy~F.~de~T\'eramond}\affiliation{\COSTARICA}
\author{Hans~G\"unter~Dosch}\affiliation{\HEIDELBERG}
\author{Stanley~J.~Brodsky}\affiliation{\SLAC}
\author{Alexandre~Deur}\affiliation{\JLAB}

\collaboration{HLFHS Collaboration}


\begin{abstract}
We propose a unified new approach to describe polarized and unpolarized quark distributions in the proton based on the gauge-gravity correspondence, light-front holography, and the generalized Veneziano model.  
We find that the spin-dependent quark distributions are uniquely determined in terms of the unpolarized distributions by chirality separation without the introduction of additional free parameters. The predictions are consistent with existing experimental data and agree with perturbative QCD constraints at large longitudinal momentum $x$. In particular, we predict the sign reversal of the polarized down-quark distribution in the proton at $x=0.8\pm0.03$, a key property of nucleon substructure which will be tested very soon in upcoming experiments.
\end{abstract}

\maketitle

{\it Introduction.}---Understanding how the spin of the proton originates from its
quark and gluon constituents is one of the most active research frontiers in hadron
physics~\cite{Deur:2018roz,Aidala:2012mv}. A key challenge is to determine the polarized parton
distribution functions (PDFs), $\Delta q(x,Q)$, which describe the difference of the
probability density between helicity-parallel and helicity-antiparallel quarks in 
a proton. Here $x$ is the light-front longitudinal momentum fraction of the proton carried by 
quarks of flavor $q$. 
The PDFs represent the universal frame-independent distribution functions of the proton which are measured in deep inelastic lepton-proton scattering (DIS) at spacelike momentum transfer $Q$.
Since they are determined by fundamental dynamics of color confinement, 
they are nonperturbative quantities.  
It is thus challenging to derive PDFs from first principles. 
However, their $x$ dependence and magnitude in the $x\to1$ limit are 
constrained by perturbative QCD (pQCD)~\cite{Farrar:1975yb, Brodsky:1994kg,Avakian:2007xa}. 
These important constraints~\cite{Farrar:1975yb,Brodsky:1994kg}, which are first-principle predictions of pQCD, predict the helicity retention at $x\sim1$; {\it i.e.}, the helicity of a quark carrying
large momentum fraction will tend to match the helicity of its parent nucleon: the helicity asymmetry $\Delta q(x,Q)/q(x,Q)$ is predicted to approach $1$ as $x\to1$,
where $q(x,Q)$ is the unpolarized PDF.

Precise measurements of $\Delta q(x,Q)$ from polarized lepton-nucleon DIS are now available~\cite{Deur:2018roz,Aidala:2012mv}.
Although the expected increase of $\Delta u/u$ toward $1$ as $x \to 1$ is observed, surprisingly $\Delta d/d$ is found to {\it remain negative} in the experimentally covered region of $x\lesssim 0.6$~\cite{Zheng:2003un,Zheng:2004ce,Parno:2014xzb,Dharmawardane:2006zd,Airapetian:2003ct,Airapetian:2004zf,Alekseev:2010ub}, 
without indication of sign reversal. Global pQCD analyses 
of the experimental data extrapolated to large $x$ favor negative values of $\Delta d/d$ at
$x\sim1$~\cite{deFlorian:2008mr,deFlorian:2009vb,Nocera:2014gqa,Jimenez-Delgado:2014xza,Ethier:2017zbq}, as do Dyson-Schwinger equation
calculations~\cite{Roberts:2013mja}. 
This contradiction with the pQCD constraint at $x\to1$ challenges our confidence in understanding the large-$x$ behavior of the polarized PDFs.

In this Letter, we present a novel unified approach to determine polarized quark distributions together with unpolarized quark distributions based on light-front holographic QCD (LFHQCD)~\cite{Brodsky:2014yha} and the generalized Veneziano model~\cite{Veneziano:1968yb,Ademollo:1969wd,Landshoff:1970ce}. 
This approach provides a natural separation of chiralities in the solution of the LFHQCD action, thus relating the matrix elements of the axial current to those of the vector current. 
Our determination of $\Delta q(x)$ 
provides an accurate description of the available experimental data and agrees with the $x\to1$ pQCD constraints.
In particular, the value of $x$ for the sign reversal of $\Delta d(x)/d(x)$ is determined, a key prediction which will be tested in upcoming experiments~\cite{E12-06-110,E12-06-122}.

LFHQCD predicts hadron structure and spectroscopy by embedding light-front dynamics in a higher dimensional gravity theory~\cite{Brodsky:2003px,Brodsky:2013ar}. This approach to color confinement provides effective semiclassical QCD bound-state equations~\cite{deTeramond:2005su,Brodsky:2006uqa,deTeramond:2008ht,Brodsky:2013ar}, where the confinement potential is determined by an underlying superconformal algebraic structure~\cite{deTeramond:2014asa,Dosch:2015nwa,Brodsky:2016yod}.
Because of the increasing interest in parton distributions, LFHQCD-based models have been developed with modifications of the light-front wave functions~\cite{Abidin:2008sb,Vega:2010ns,Vega:2012iz,Gutsche:2013zia,Gutsche:2014yea,Chakrabarti:2013gra,Chakrabarti:2015ama,Dehghani:2015jva,Chakrabarti:2017teq,Mondal:2016xsm,Maji:2016yqo,Maji:2017bcz,Traini:2016jru,Gutsche:2016gcd,Maji:2017ill,Rinaldi:2017roc,Bacchetta:2017vzh,Mondal:2017wbf,Chouika:2017dhe,Xu:2019xhk}. 
Recently, we introduced a new approach for deriving PDFs as well as generalized parton distributions (GPDs) from LFHQCD~\cite{deTeramond:2018ecg} with minimal parameters. 
The nucleon and pion unpolarized PDFs were simultaneously determined using the universality properties of parton distributions in LFHQCD~\cite{deTeramond:2018ecg}. The pion PDF at large $x$ was predicted to decrease as $(1-x)$, while pQCD~\cite{Farrar:1975yb} and Dyson-Schwinger equation results~\cite{Chen:2016sno} suggest $(1-x)^2$. While for the pion distribution an even power of $(1-x)$ is required by the Gribov-Lipatov reciprocity relation~\cite{Gribov:1972ri,Gribov:1972rt} based on the $t\to s$ sign-changing crossing behavior of perturbative amplitudes, the power of $(1-x)$ for a nonperturbative confining interaction is still unsettled.

Motivated by the encouraging results of Ref.~\cite{deTeramond:2018ecg}, we extend the formalism here to polarized distributions. As spin has a notorious history of invalidating theoretical predictions, the exploration of spin-dependent observables is a crucial test. Using the chirality separation from the LFHQCD action, we find that the unpolarized and polarized quark distributions are strictly related, which allows us to predict polarized distributions from unpolarized ones without adding free parameters.

{\it Formalism.}---We first review the derivation of unpolarized proton PDFs from the holographic expression of its spin-non-flip Dirac form factor $F_1(t)$, where $t=-Q^2$ is the square of transferred momentum. 
The contribution from a twist-$\tau$ Fock state in the light-front Fock expansion of the proton wave function, a component with effectively $\tau$ constituents, to the Dirac form factor is given by~\cite{Brodsky:2014yha,Brodsky:2008pg}
\begin{align}
    F_{1}(t) = c_{V,\tau} F_{V,\tau}(t) + c_{V,\tau+1}F_{V,\tau+1}(t),
    \label{eq:F1}
\end{align}
with
\begin{align}
    F_{V,\tau}(t) = \frac{1}{N_{V,\tau}} B\Big(\tau-1, \frac{1}{2}-\frac{t}{4\lambda}\Big).
    \label{eq:FVtau}
\end{align}
The subscript $V$ indicates the coupling to a vector current and $\lambda=\kappa^2$ is the universal mass scale in LFHQCD which can be fixed by hadron spectroscopy; the fit to the $\rho/\omega$ trajectory gives $\kappa =0.534\,\rm GeV$. The $c_{V,\tau}$ and $c_{V,\tau+1}$ are coefficients to be determined, $N_{V,\tau}$ is a normalization factor, and $B(x,y)$
is the Euler Beta function.
The two terms in Eq.~\eqref{eq:F1} correspond respectively to the contribution from the {\it even} and {\it odd} chiral components, $\Psi_+$ and $\Psi_-$, of the bulk field solution~\cite{Brodsky:2014yha}, providing a natural chirality separation. 
Equation~\eqref{eq:FVtau} has the same structure as a generalization of the Veneziano amplitude $B\big(1-\alpha(s),1-\alpha(t)\big)$~\cite{Veneziano:1968yb} to nonstrong process~\cite{Ademollo:1969wd,Landshoff:1970ce}, here electron-nucleon scattering. 
Our framework thus incorporates nonperturbative analytic structures found in pre-QCD studies, such as Regge trajectories and generalized Veneziano amplitudes. 

The $t$ dependence in Eq.~\eqref{eq:FVtau} can be rewritten as $1-\alpha_V(t)$ with the Regge trajectory~\cite{deTeramond:2018ecg}
\begin{align}
    \alpha_V(t) = \frac{t}{4\lambda} + \frac{1}{2}.
    \label{eq:alphaV}
\end{align}
This is just the $\rho/\omega$ trajectory emerging from LFHQCD for vector mesons with massless quarks~\cite{Brodsky:2016yod}. The quark mass correction is negligible for $u$ and $d$ quarks; for the strange quark contribution, the $\phi$ trajectory shifts the intercept to $\alpha_{\phi}(0)\approx0.01$~\cite{Sufian:2018cpj}. 

With the integral representation of the Beta function, the unpolarized PDF for a twist-$\tau$ state is
\begin{align}
    q(x) = c_{V,\tau} q_\tau(x) + c_{V,\tau+1} q_{\tau+1}(x),
    \label{eq:unpolPDF}
\end{align}
where
\begin{align}
    q_\tau(x) &= \frac{1}{N_{V,\tau}}w(x)^{-\frac{1}{2}}[1-w(x)]^{\tau-2}w'(x),
\end{align}
and $w(x)$ is a universal reparametrization function highly constrained by several boundary conditions~\cite{deTeramond:2018ecg}.

Now, we turn to the polarized distributions, for which the coupling of an axial current
--rather than a vector current--is needed. Since the current operator differs by a $\gamma_5$, the axial form factor follows Eq.~\eqref{eq:F1}, but with a sign flip from the contribution of the chiral-odd component,
\begin{align}
    F_{A}(t) = c_{A,\tau} F_{A,\tau}(t) - c_{A,\tau+1} F_{A,\tau+1}(t),
    \label{eq:FA}
\end{align}
where
\begin{align}
    F_{A,\tau}(t) = \frac{1}{N_{A,\tau}} B\Big(\tau - 1, 1 - \frac{t}{4\lambda}\Big),
    \label{eq:FAtau}
\end{align}
and the subscript $A$ indicates the coupling to an axial current. $F_{A,\tau}(t)$ has the same  structure as $F_{V,\tau}(t)$, but with the Regge trajectory replaced by the axial one,
\begin{align}
    \alpha_{A}(t) = \frac{t}{4\lambda},
    \label{eq:alphaA}
\end{align}
emerging from LFHQCD~\cite{Brodsky:2016yod}. The coefficients in Eq.~\eqref{eq:FA} and those in Eq.~\eqref{eq:F1} are related since they correspond to the same state. Thus apart from the sign flip in the second term in Eq.~\eqref{eq:FA}, they have the same value relative to the normalization factors as given by
\begin{align}
    \frac{c_{V,\tau}}{N_{V,\tau}} = \frac{c_{A,\tau}}{N_{A,\tau}}.
    \label{eq:normrelation}
\end{align}
Since the normalization convention is arbitrary, we set $N_{V,\tau}=N_{A,\tau}=N_\tau$, and therefore  identify the coefficients as $c_{V,\tau}=c_{A,\tau}=c_{\tau}$~\cite{note1}.

Following the same procedure, we express the $\Delta q(x)$ for a twist-$\tau$ state as
\begin{align}
    \Delta q(x) = c_\tau \Delta q_\tau (x) - c_{\tau+1} \Delta q_{\tau+1}(x),
    \label{eq:polPDF}
\end{align}
where
\begin{align}
    \Delta q_\tau(x) = \frac{1}{N_\tau} [1 - w(x)]^{\tau-2} w'(x).
\end{align}

At large $x$, we expand $w(x)$ near $x=1$~\cite{deTeramond:2018ecg},
\begin{align}
    w(x) = 1 + \frac{1}{2}w''(1)(1-x)^2 + {\cal O}\big((1-x)^3\big),
\end{align}
and find that $q_\tau(x)$ and $\Delta q_\tau(x)$ have the same behavior,
\begin{align}
    q_\tau(x) = \Delta q_\tau(x) = \frac{[-w''(1)]^{\tau-1}}{2^{\tau-2} N_\tau}(1-x)^{2\tau-3}
    +\cdots,\label{eq:largex}
\end{align}
where higher powers of $(1-x)$ are suppressed. For both $q(x)$, Eq.~\eqref{eq:unpolPDF}, and $\Delta q(x)$, Eq.~\eqref{eq:polPDF}, the function is dominated by the first term at large $x$, unless its coefficient $c_\tau$ vanishes.
Then the helicity asymmetry at $x\to1$ is
\begin{align} \label{ha}
    \lim_{x\to1} \frac{\Delta q(x)}{q(x)} = 1,
\end{align}
consistent with the pQCD constraint~\cite{Farrar:1975yb,Brodsky:1994kg}.

The spin-aligned and spin-antialigned distributions are linear combinations of the unpolarized and polarized distributions:
\begin{align}
    q_\uparrow(x) &= \frac{1}{2}[q(x) + \Delta q(x)],\\
    q_\downarrow(x) &= \frac{1}{2}[q(x) - \Delta q(x)].
\end{align}
We find, in the large-$x$ limit,
\begin{align}
    q_\uparrow(x) &\rightarrow c_\tau q_{\tau}(x),\label{eq:quparrowlim}\\
    q_\downarrow(x) &\rightarrow c_{\tau+1} q_{\tau+1}(x).\label{eq:qdownarrowlim}
\end{align}
The spin-aligned or spin-antialigned  distributions tend, respectively, to a pure contribution from the chiral-even or chiral-odd component, $\Psi_+$ or $\Psi_-$, of the nucleon bulk field solution, consistent with pQCD helicity retention at large $x$. Equations~\eqref{eq:quparrowlim} and~\eqref{eq:qdownarrowlim} provide the asymptotic normalization, which can be used to derive the same relation as in Eq.~\eqref{eq:normrelation}. 

From Eq.~\eqref{eq:largex}, $q_\uparrow(x)$ and $q_\downarrow(x)$ decrease as $(1-x)^{2\tau-3}$ and $(1-x)^{2\tau-1}$, respectively. For the valence state $\tau=3$, they behave as $(1-x)^3$ and $(1-x)^5$, consistent with pQCD~\cite{Farrar:1975yb}, up to logarithmic corrections~\cite{Brodsky:1994kg,Avakian:2007xa}.

At small $x$, $w(x)\sim x$. Thus $\Delta q(x)$ decreases faster than $q(x)$ with decreasing $x$, and the helicity asymmetry behaves as
\begin{align}
    \frac{\Delta q(x)}{q(x)} \sim x^{\frac{1}{2}},
\end{align}
where the exponent $\sfrac{1}{2}$ is given by the difference between the intercepts of the vector and axial Regge trajectories \eqref{eq:alphaV} and \eqref{eq:alphaA}; the intercepts are shifted by a negligible amount when $u$ and $d$ quarks mass corrections are included. When $x\to0$, the helicity asymmetry goes to zero,
\begin{align} 
\lim_{x\to0}\frac{\Delta q(x)}{q(x)} = 0,
\end{align}
which indicates that the helicity correlation between a quark and its parent nucleon disappears. This result is natural~\cite{Brodsky:1994kg} because the constituents and the nucleon have infinite relative rapidity for $x\sim 0$. This property is confirmed by the experimental data~\cite{Nocera:2014uea}.
    
{\it Numerical results.}---Up to now, all results have been derived for arbitrary twist-$\tau$ components without any specific choices for $c_\tau$ or for $w(x)$, as long as the general boundary conditions are fulfilled. If only valence states are considered, we can express the flavor Dirac form factors of $u$ and $d$ quarks as
\begin{align}
    F_1^u(t) &= c_{3,u} F_{V,3}(t) + (2-c_{3,u}) F_{V,4}(t),\label{eq:F1uv}\\
    F_1^d(t) &= c_{3,d} F_{V,3}(t) + (1-c_{3,d}) F_{V,4}(t),\label{eq:F1dv}
\end{align}
where the quark number sum rule has been applied, with $N_\tau=B(\tau-1,1/2)$ normalizing $F_{V,\tau}(0)$ to $1$. The sea quark constituents, beyond the valence state, are encoded in 
higher Fock states with additional quark-antiquark pairs. 
In this work, we will truncate the Fock expansion of the nucleon state up to one quark-antiquark pair, which is a twist-$5$ state.
As a simplifying procedure to include the sea quark contributions we can add to Eqs.~\eqref{eq:F1uv} and~\eqref{eq:F1dv} the term
\begin{align}
    c_{5,u/d} F_{V,5}(t) - c_{5,u/d} F_{V,6}(t),
\end{align}
which assumes that the quark number sum rule is saturated by the contribution from the valence quarks.  
One can also include the intrinsic strange contribution as in Ref.~\cite{Sufian:2018cpj}. We will compare the three cases: i) only the valence state contribution; ii) including the contribution from the $u\bar{u}$ and $d\bar{d}$ pairs; iii) also including the contribution from the intrinsic strange sea,  using the holographic results from our previous work~\cite{Sufian:2018cpj}. We fix the coefficients by matching to the Dirac form factor~\cite{Ye:2017gyb}, as listed in Table~\ref{tab:parameters}.

\begin{table}[htp]
    \centering
    \caption{Coefficient values fixed by matching to the electromagnetic form factor~\cite{Ye:2017gyb}. The parameter $a$ in Eq.~\eqref{eq:wxform} is fixed by the first moment of the unpolarized valence quark distributions. The last column $g_{A,\rm min}$ is the isovector axial charge with minimal sea. The meaning of the labels is explained in the text.}
    \begin{tabular}{m{0.1\columnwidth}m{0.1\columnwidth}m{0.1\columnwidth}m{0.12\columnwidth}m{0.12\columnwidth}>{\centering}m{0.1\columnwidth}m{0.11\columnwidth}m{0.11\columnwidth}}
    \hline\hline
    Label  & $c_{3,u}$ & $c_{3,d}$ & $c_{5,u}$ & $c_{5,d}$ & $a$ & $g_{A,\rm min}$ & $\delta_{u/d}^{\rm max}$\\
    \hline
    I & 1.782 & 0.066 & ... & ... & 0.407 & 0.867 & 0.342 \\
    II & 1.793 & 0.062 & $-0.559$ & $-0.516$ & 0.480 & 0.879 & 0.332 \\
    III & 1.794 & 0.060 & $-0.492$ & $-0.447$ & 0.471 & 0.881 & 0.330\\
    \hline\hline
    \end{tabular}
    \label{tab:parameters}
\end{table}

Since the electromagnetic form factors only measure the difference between quark and antiquark contributions,
namely, $c_{\tau,u} \equiv u_\tau - \bar u_\tau$ and similarly for the $d$ quark, contributions to $u_\tau$ and $\bar{u}_\tau$ cannot be uniquely separated.
However, a lower limit can be derived
from the positivity bounds $q_\uparrow(x)\geq0$ and $q_\downarrow(x)\geq0$. With the asymptotic relations~\eqref{eq:quparrowlim} and~\eqref{eq:qdownarrowlim}, this requirement is fulfilled by the minimal sea contribution,
\begin{align}
    \bar{u}(x)_{\rm min} &= c_{5,u} q_{\tau=6}(x)\quad {\rm if}\quad c_{5,u}\geq0,\\
    \bar{u}(x)_{\rm min} &= -c_{5,u} q_{\tau=5}(x)\quad {\rm if}\quad c_{5,u}<0,
\end{align}
and similarly for $\bar{d}$. This constraint is stronger than that utilized in Ref.~\cite{Sufian:2018cpj}, where only the sum $q_\uparrow(x)+q_\downarrow(x)\geq0$ is required.

Since the sea quark distributions are not separately constrained by the electromagnetic form factors, one needs other physical observables that are sensitive to the quark and antiquark contributions individually to determine them separately. Instead of attempting a full separation, which is beyond the purpose of this Letter, we use the relation of the isovector axial charge,
\begin{align}
    g_A = (\Delta u + \Delta \bar{u}) - (\Delta d + \Delta \bar{d}),
    \label{eq:gAsumrule}
\end{align}
to constrain the nonminimal sea quark.

The value of the isovector axial charge is precisely determined by the neutron weak decay~\cite{Tanabashi:2018oca}, $g_A=1.2732(23)$.  As shown in Table~\ref{tab:parameters}, its value evaluated with a minimal sea component, $g_{A,\rm min}$, is smaller.  
To obtain the value of $g_A$ with the minimal shift $u_\tau \to u_\tau + \delta_{\tau,u}$,  $\bar u_\tau \to \bar u_\tau + \delta_{\tau,u}$ and similarly for the $d$ quark, implies a positive shift $\delta_{\tau=5,u}$ and/or $\delta_{\tau=6,d}$. Therefore, we satisfy the relation~\eqref{eq:gAsumrule} by the shift  $\delta_{\tau=5,u}$ and $\delta_{\tau=6,d}$, and take the variation between them as part of the theoretical uncertainty. The maximal ranges, $\delta_{u/d}^{\rm max}$, of this shift are listed in Table~\ref{tab:parameters}.

\begin{figure}[htp]
    \centering
    \includegraphics[width=\columnwidth]{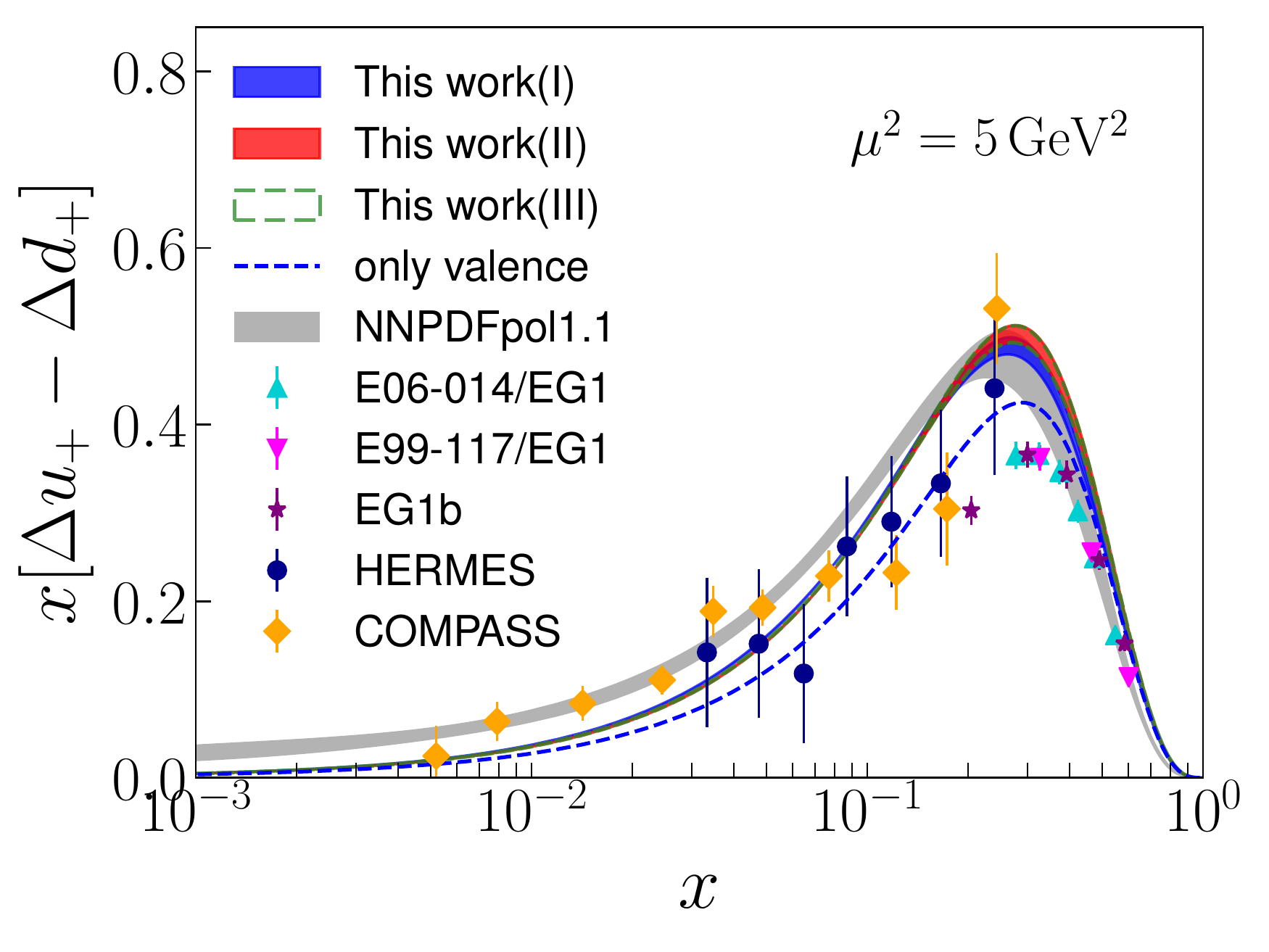}
    \caption{Polarized distributions of the isovector combination $x[\Delta u_+(x)-\Delta d_+(x)]$ in comparison with NNPDF global fit~\cite{Nocera:2014gqa} and experimental data~\cite{Zheng:2003un,Zheng:2004ce,Parno:2014xzb,Dharmawardane:2006zd,Airapetian:2003ct,Airapetian:2004zf,Alekseev:2010ub}. Three sets of parameters, see Table~\ref{tab:parameters}, are determined from the Dirac form factor and unpolarized valence distributions. The bands represent the variation with different approaches to saturate the axial charge isovector relation~\eqref{eq:gAsumrule}. The blue dashed curve is the valence state contribution without saturating the axial charge.}
    \label{fig:pol-uub-ddb}
\end{figure}

\begin{figure}[htp]
    \centering
    \includegraphics[width=\columnwidth]{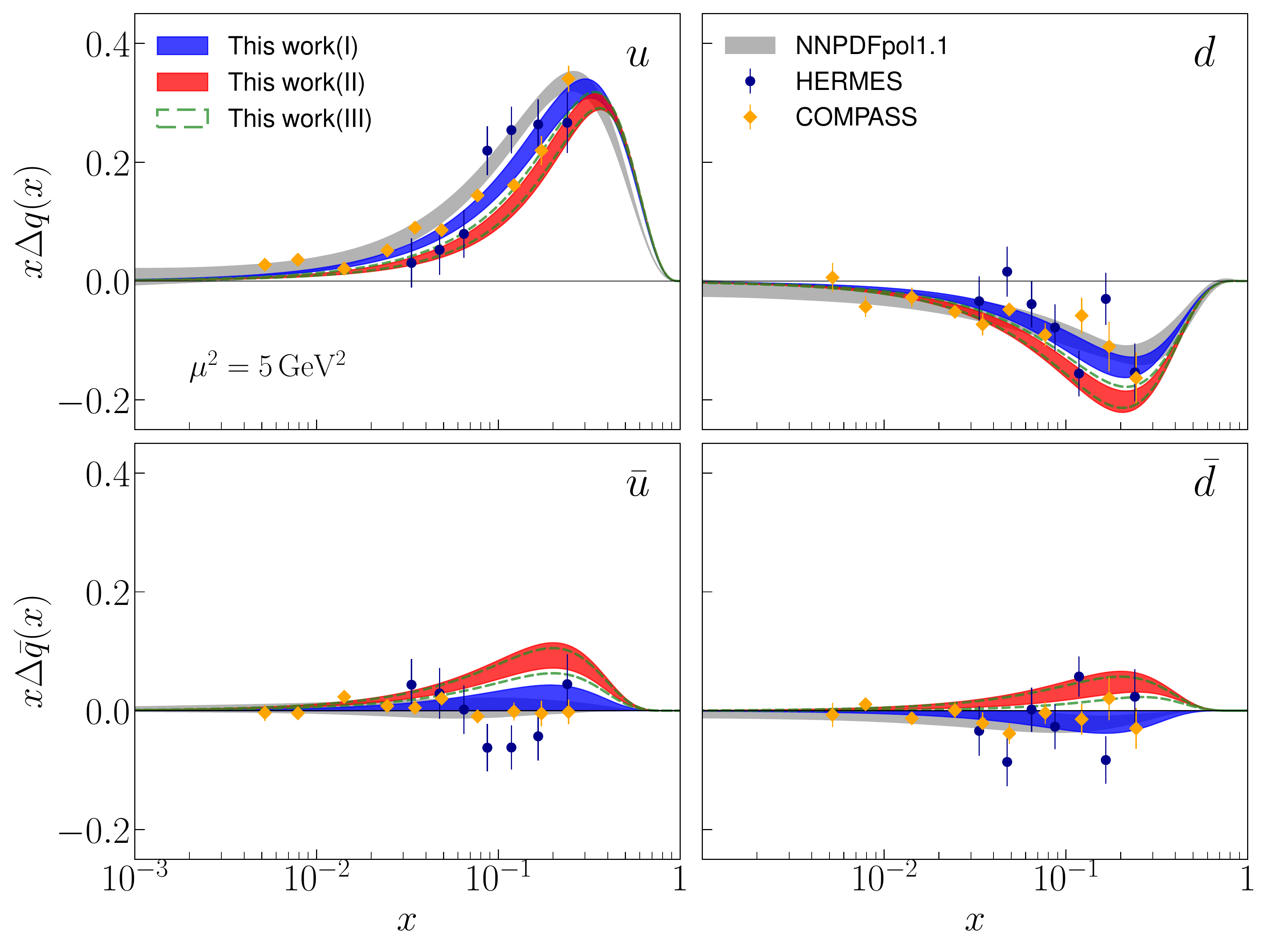}
    \caption{Polarized distributions of $u$, $d$, $\bar{u}$, and $\bar{d}$. 
    The bands and symbols have the same meaning as in Fig.~\ref{fig:pol-uub-ddb}.}
    \label{fig:polq}
\end{figure}

\begin{figure}[htp]
    \centering
    \includegraphics[width=\columnwidth]{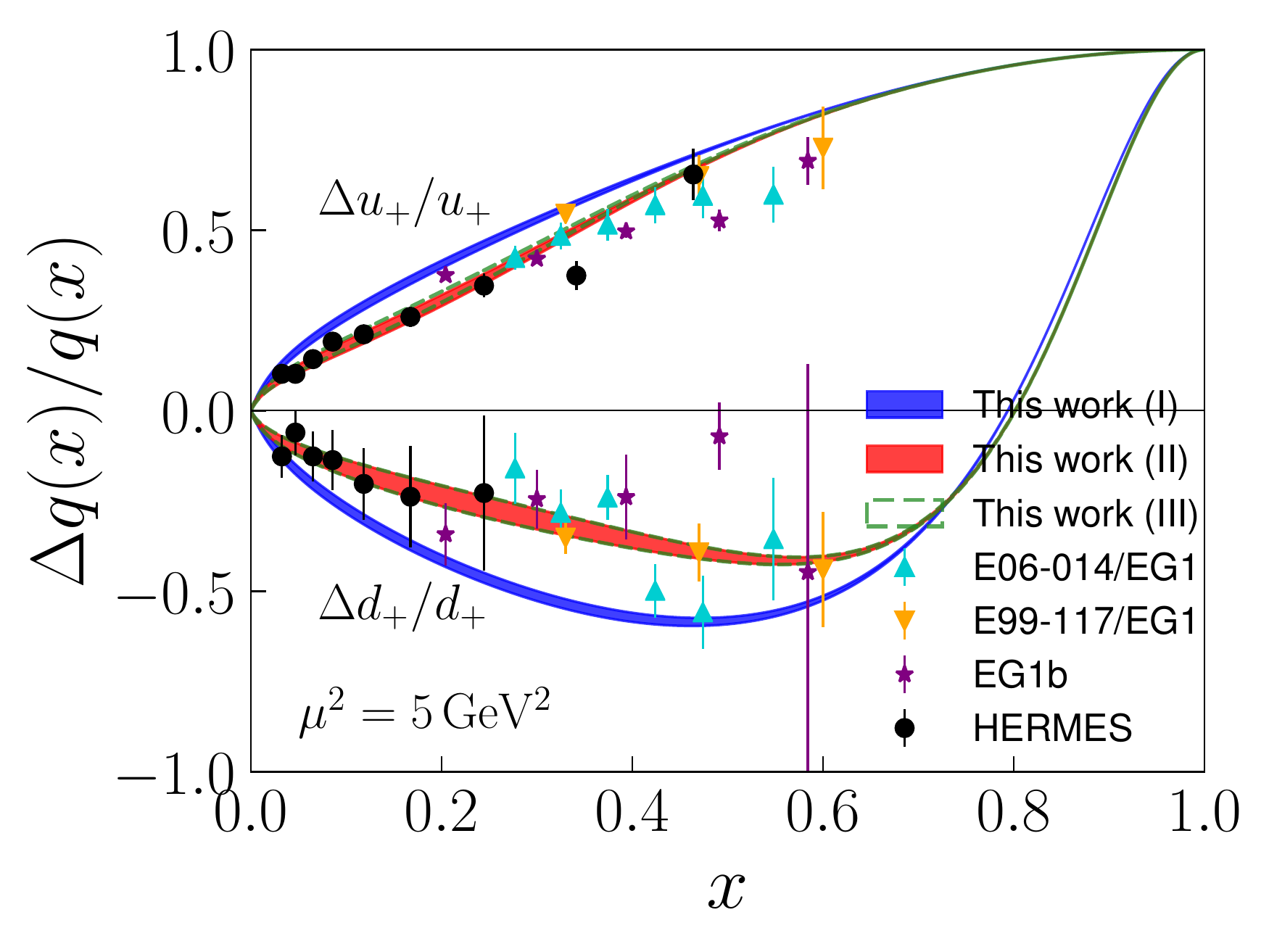}
    \caption{Helicity asymmetries of $u+\bar{u}$ and $d+\bar{d}$. The bands and symbols have the same meaning as in Fig.~\ref{fig:pol-uub-ddb}.}
    \label{fig:dqoverq}
\end{figure}

For the universal reparametrization function $w(x)$, we take the same form as in Ref.~\cite{deTeramond:2018ecg},
\begin{align}
    w(x) = x^{1-x}\exp[-a(1-x)^2],
    \label{eq:wxform}
\end{align}
with the parameter $a$ fixed with the first moment of unpolarized valence quark distributions. One can, in principle, adopt any parametrization form that fulfills the boundary conditions specified in Ref.~\cite{deTeramond:2018ecg}. 
The predictive power is kept by the universality of $w(x)$ for all PDFs. For comparison with measurements, we evolve the PDFs from $Q=1.06\,\rm GeV$, which is the matching scale determined from the study of the strong coupling constant~\cite{Deur:2016opc}, to $Q=\sqrt{5}\,\rm GeV$. 
As shown in Figs.~\ref{fig:pol-uub-ddb}-\ref{fig:dqoverq}, our numerical results agree with the global fit~\cite{Nocera:2014gqa} and measurements~\cite{Zheng:2003un,Zheng:2004ce,Parno:2014xzb,Dharmawardane:2006zd,Airapetian:2003ct,Airapetian:2004zf,Alekseev:2010ub}. The isovector combination $\Delta u_+ - \Delta d_+$, where $u_+$ and $d_+$ stand for $u+\bar{u}$ and $d+\bar{d}$, is the distribution relevant to the relation of the axial charge~\eqref{eq:gAsumrule}. 
The dashed blue curve in Fig.~\ref{fig:pol-uub-ddb} is the contribution from the valence state only;
the difference with the full results, cases I, II, and III, which include saturation of the axial charge, is noticeable. This is consistent with the analysis of the Pauli form factor in Ref.~\cite{Sufian:2016hwn}, which demonstrates the significance of the sea quarks in describing spin-related quantities. 

As shown in Fig.~\ref{fig:polq}, the variation of our predictions for each flavor from the three sets of coefficients is large, since the sea quark coefficients are not well constrained by the procedure discussed above. Furthermore, the truncation of the Fock state up to five-quark states, which allows only one pair of sea quarks, may potentially result in greater theoretical uncertainties for each individual flavor. Equation~\eqref{eq:gAsumrule} provides an important constraint, but it still leaves some flexibility, such as adding the same term to $u\bar{u}$ and $d\bar{d}$. Since the goal of this Letter is to introduce a new approach for the study of polarized PDFs, we leave this issue to more detailed future investigations. 

Most important, the critical region for the upcoming Jefferson Lab
spin program~\cite{E12-06-110,E12-06-122} is the large-$x$ region, which is dominated by the valence state and is thus much less affected by the variation of the sea. As observed in Fig.~\ref{fig:dqoverq}, the predictions with three sets of coefficients are consistent, especially in the large-$x$ region. As we have analytically demonstrated above, our approach supports the pQCD prediction that the helicity asymmetry (\ref{ha}) approaches $1$ in the large-$x$ limit following the power behavior $(1-x)^2$, the ratio of spin-antialigned to spin-aligned distributions~\cite{Farrar:1975yb}. In particular, the sign reversal of the $d$-quark helicity distribution in the proton is robustly predicted to be $x=0.80\pm0.03$, where the theoretical uncertainty includes the variation among the three sets of coefficients in Table~\ref{tab:parameters} and the scale $Q$ from $\sqrt{3}$ to $\sqrt{5}$\,GeV. 

{\it Summary.}---We have presented a new unified approach to describe spin-dependent and spin-independent quark distributions from nonperturbative confining dynamics  based on the gauge-gravity correspondence, light-front holography and the generalized Veneziano model. 
The chirality separation property of the holographic form factor structures strictly relates polarized and unpolarized quark distributions. With all parameters fixed by the nucleon Dirac form factor and unpolarized quark distributions, our predictions for the polarized distributions agree with existing data. Our analytic results for $\Delta q(x)/q(x)$ are consistent with the large-$x$ behavior predicted by pQCD. Our analysis also supports the pQCD prediction of helicity retention at $x\sim1$; this fundamental prediction has been challenged by Dyson-Schwinger equation calculations, but it has not yet been constrained by data.
In the large-$x$ region, where the valence state dominates, we predict that the $d$-quark helicity will flip sign at $x=0.80\pm0.03$, regardless of the procedure used to include the sea quark contribution. This prediction will be tested soon~\cite{E12-06-110,E12-06-122}. The analytic behavior at large $x$ and the agreement with existing data reinforces confidence in the pQCD prediction, which can be implemented in global analysis such as that of Refs.~\cite{Leader:2001kh,Bourrely:2001du}. In addition, the relation between the unpolarized and polarized distributions can be tested by simultaneous fits to unpolarized and polarized PDFs.

\acknowledgments{
We would like to thank J.-P.~Chen for helpful discussions. This work is supported in part by the U.S. Department of Energy, Office of Science, Office of Nuclear Physics under contracts No. DE-AC05-06OR23177 and No. DE-FG02-03ER41231. This work is also supported in part by National
Natural Science Foundation of China under Contract No. 11775118.
}


\end{document}